# Coupling Physics in Machine Learning to Predict Properties of High-temperatures Alloys

Jian Peng[a], Yukinori Yamamoto[a], Jeffrey A. Hawk[b], Edgar Lara-Curzio[a], Dongwon Shin[a,*]

[a] *Materials Science and Technology Division, Oak Ridge National Laboratory, Oak Ridge, TN 37831*

[b] *Materials Performance Division, National Energy Technology Laboratory, Albany, OR 97321-2198*

* Email address: shind@ornl.gov

**Abstract:** High-temperature alloy design requires a concurrent consideration of multiple mechanisms at different length scales. We propose a workflow that couples highly relevant physics into machine learning (ML) to predict properties of complex high-temperature alloys with an example of the 9-12 wt.% Cr steels yield strength. We have incorporated synthetic alloy features that capture microstructure and phase transformations into the dataset. Identified high impact features that affect yield strength of 9Cr from correlation analysis agree well with the generally accepted strengthening mechanism. As part of the verification process, the consistency of sub-datasets has been extensively evaluated with respect to temperature and then refined for the boundary conditions of trained ML models. The predicted yield strength of 9Cr steels using the ML models is in excellent agreement with experiments. The current approach introduces physically meaningful constraints in interrogating the trained ML models to predict properties of hypothetical alloys when applied to data-driven materials.

**Introduction**

Material design assisted by data analytics is an emerging area of materials science and engineering that offers a reduction in cost, risk, and time over traditional material development approaches based solely on experimental investigations and/or physics-based simulations[1-5]. Due to their complexity (i.e., chemistry, melt process, thermo-mechanical process, heat treatment, and resulting developed microstructure), the rational design of high-temperature alloys by machine learning (ML) requires a comprehensive dataset that can cover various aspects: multi-component, multi-phase, multi-physics, multi-scale and multiple strengthening mechanisms as well as significant influence of processing conditions on the properties of final products.

The majority of previous efforts applying ML to predict the properties of high-temperature alloys have used alloy compositions and processing conditions as features[6-13]. While these approaches can leverage experimental data accumulated over decades, extrapolating these models outside the range of the input data is risky due to the absence of physical constraints. There have been attempts to incorporate atomistic-level features, e.g., atomic radius/volume, electronegativities, cohesive energy, and local electronegativity mismatch, for predicting high-temperature alloy properties[14-17], but features related to phenomena/mechanisms occurring in larger length-scales (i.e., micro- and meso-scale) may have more impact on alloys.

For high-temperature alloy design, physical information, such as microstructure, is essential for representing process-structure-property correlation[18-22]. Pioneering work by Zhao and Henry showed that the performance of a regression model for predicting the rupture time of Ni-based alloys could be significantly improved by incorporating the equilibrium volume fraction of the γ´ phase[21]. Recently, further advancement in this area was made by establishing a data analytics workflow by integrating microstructure-related synthetic features via the CALPHAD approach to predict the creep strength of alumina-forming austenitic stainless steels[23] and high-strength stainless steel[24].

However, for many material systems (high-temperature alloys in particular), microstructure-related synthetic features from CALPHAD are often not enough since the microstructure changes over time and the strengthening mechanisms evolve with applied stress and temperature. Consider the case of 9-12 wt.% Cr martensitic-ferritic steels (hereafter referred to as 9Cr steel) as an example. This class of alloy consists of a tempered martensitic microstructure, where temperature plays a critical role with respect to strengthening mechanisms[25]. Fine prior austenite grains/packet/lath structure and dislocation density significantly control the room-temperature strength. With increasing temperatures, up to around 600~650°C, second-phase precipitates, i.e., $M_{23}C_6$ (M=Fe, Cr, and Mn), MX (M: mainly V, X: C and N), and even Laves phase, within the sub-grain interior or along these sub-boundary, play an important role in strengthening. Above 700°C, microstructure instability, such as rapid precipitate coarsening, recovery and/or recrystallization, lead to a significant loss in mechanical strength.

Thus, relevant features such as phase transformation temperatures, e.g., A3 temperature (the temperature at which transformation of ferrite to austenite is completed during heating) and martensite start temperature (Ms) etc., should be considered, in addition to microstructure information. These phase transformation temperatures are directly correlated with martensitic microstructure evolution, other microstructural features (e.g., prior austenite grain size, packet/lath sizes, dislocation density in the as-normalized condition, etc.[26,27]), and consequently influence their initial mechanical properties as well as long-term microstructural stability.

Herein, we demonstrate a workflow of coupling highly relevant physics into ML models for predicting properties of multi-phase and multi-component high-temperature alloys. A yield strength dataset of 9Cr martensitic-ferritic steels is selected to elucidate this strategy. Figure 1 illustrates the structure of the yield strength dataset of 9Cr steel used in this study. The computed synthetic alloy features, along with raw experimental data, are listed in Table 1. The correlation between these features in the dataset and the 9Cr yield strength was quantitatively determined and compared with generally accepted mechanisms by the community. We evaluated the performance of representative ML models, i.e., linear regression (LR)[28], Bayesian ridge (BR)[29,30], $k$-nearest neighbor (NN)[31], random forest (RF)[32], and support vector machines (SVM)[33]. Additional work was also carried out to assess the performance of ML models on predicting the prior austenite grain size (PAGS) of the 9Cr steel since PAGS is an essential input for calculating Ms of 9Cr steels.

[Table 1 about here]

[Figure 1 about here]

**Results and Discussion**

We started with the 9Cr dataset with only raw experimental data (i.e., elemental alloy compositions, processing, and testing conditions, and PAGS – groups 1 and 2 in Figure 1) to train five different ML models. Figure 2 shows the average accuracy of these models, and their standard deviation from 10 training runs as a function of the numbers of top-ranking features from Pearson's correlation coefficient (PCC)[34] and maximal information coefficient (MIC)[35] analyses. Overall, ML models RF, NN, and SVM exhibit high accuracy ($R^2>0.9$) regardless of the number of top-ranking features. More specifically, RF was the most accurate (always higher than 0.95), followed by SVM. Nevertheless, the applicability of these models for alloy design is questionable since PAGS is the only physically measured microstructure related feature involved in ML training. Other relevant physically meaningful features, such as volume fraction of key phases and phase transformation temperatures, are required to properly represent the process-structure-property relationship and serve as physical constraints in ML.

[Figure 2 about here]

*Analyses of temperature-based sub-datasets*

Given the lack of physically measurable microstructure features in the 9Cr dataset, the raw experimental data were augmented with synthetically derived features, i.e., groups 3 and 4 in Figure 1 (see Table 1), from high-throughput CALPHAD calculations. Since the primary strengthening mechanisms of 9Cr steel are temperature-dependent, it was essential to carefully examine whether the present dataset is capable of representing the temperature-dependent strengthening mechanism. Thus, we divided the 9Cr dataset into several sub-datasets based on the testing temperature for further analysis. As such, we performed correlation analysis for each sub-dataset.

The top 10 and bottom 10 features from the PCC analysis were evaluated at three representative temperatures, i.e., 200°C (low temperature), 550°C and 650°C (medium to high temperatures), and 750°C (above service temperature). These results are presented in Figure 3. From this analysis, it was observed that the closer the absolute value of the correlation coefficient is to 1, the stronger the correlation is between the feature and yield strength. Those features identified with either a positive or negative correlation with yield strength at 200, 550, and 650°C were consistent and mostly in good agreement with generally accepted strengthening factors/mechanisms in 9Cr steel. For example, Ni content exhibited a strong positive correlation with yield strength, i.e., the higher the Ni content, the higher the yield strength. This is in accordance with the practice of adding Ni to 9Cr steel to stabilize austenite at high temperatures, lower the martensitic transformation temperatures, and consequently, increases the hardenability in the normalization process. These effects generally increase the yield strength of martensitic-ferritic steels, including the 9Cr family of steel[27]. This result is also logistically supported by the present correlation analysis that shows a strong negative correlation between the martensitic start temperatures (Ms) and the yield strength.

The $M_{23}C_6$ phase also plays an important role in strengthening the 9Cr steel from the precipitate strengthening perspective and stabilizes the tempered martensite microstructure, especially at elevated temperatures[36]. A higher volume fraction of $M_{23}C_6$ leads to higher yield strength. Thus, it is reasonable that the volume fraction of $M_{23}C_6$ has one of the strongest positive correlations with yield strength. The elements V and N facilitate the formation of strengthening MX precipitates during tempering, which also assists in increasing yield strength by impeding dislocation motion during deformation and stabilizing the sub-grain structure. Co is also an austenite stabilizer that suppresses $\delta$-ferrite formation during the normalizing heat treatment step. Ms and microstructure-related features (e.g., volume fractions of $M_{23}C_6$, hcp, and fcc phases) from our high-throughput calculation are highly impactful features, critical to obtaining high-fidelity surrogate ML models. This finding is also applicable to the other sub-datasets up to 650°C (see Supplementary Table S1).

For the sub-datasets above 650°C (for example, 750°C in Figure 3), the correlation coefficients are smaller than those at low temperatures, indicating weaker response between alloy features and yield strength. In addition, the feature ranking order at 750°C is counterintuitive and very different from the trends below 650°C. For instance, Ms has a negative correlation below 650°C, and now it shows to have a positive response at 750°C. Features wC, wCr, wW, and PAGS should positively contribute to yield strength are now identified as having a negative impact at 750°C. The MIC analysis also shows a similar trend (see Supplementary Figure S1).

The correlation between alloy features and yield strength at 750°C is much weaker than those at lower temperatures. Typical high impact features, such as Temper1, wV, wNb, wNi, wC, T2_VPV_M23C6, have been correctly identified at 200, 550, and 650°C, while at 750 °C they are counterintuitive in nature. The present findings may be put into context by realizing that (i) the number of data points at >650°C is insufficient for representing the effects of certain features on yield strength correctly, and (ii) the microstructure changes during exposure at high temperatures are significant and may result in a variation of yield strength attributed to other factors that are not considered in the present dataset (e.g., the heating rate and/or the holding time before tensile testing at temperature).

[Figure 3 about here]

We then trained five ML models (BR, LR, RF, NN, and SVM) with these sub-datasets at each temperature. Since these sub-datasets have a maximum of 44 data points, we limited the number of top-ranking features used in machine learning to 10 to avoid overfitting. The top 10 features of each sub-dataset from correlation analysis are summarized in the supplementary information (Table S1). As an example, Figure 4 shows the accuracy of the RF model trained with various top-ranking features as a function of temperature-based sub-datasets. The results of the 9Cr entire dataset ("All") are also included for comparison. As shown in Figure 4a, the accuracy of ML models trained with sub-datasets is always lower than that of the one using the entirety of the 9Cr dataset (i.e., "All"), which can be attributed to their smaller volume of data for the former. The performance of RF trained with top-ranking features from MIC does not improve with more features, and the top-4 features already lead to the maximal accuracy. This exercise shows that these features are sufficient to fit the RF model well. However, the top-8 features from PCC analysis are required to reach maximal accuracy (Figure 4b). In both cases, the maximum accuracy is always greater than 0.8 from room temperature (RT) to 600°C regardless of the ranked features from the MIC or PCC analyses. From this point, it decreases monotonically above 600°C, which is in accordance with the decreasing data volume above 600°C (see Figure 1). Since the ranking of features at 650°C is reasonable (see Figure 3), the lower accuracy at 650°C may be attributed to its slightly smaller data volume than the lower temperature datasets.

Consequently, no matter how many top-ranking features are used in ML models, the accuracy ($R^2$) is always below 0.6. This observation again confirms that data at >650°C are insufficient, and the features in the present 9Cr dataset cannot represent the microstructure instability at high temperatures. Therefore, including the data at >650°C could mislead the training of ML models, and consequently, result in an incorrect prediction. For this reason, data above 650°C were removed, resulting in the truncated (≤650°C) 9Cr dataset for the following ML model.

[Figure 4 about here]

*Truncated (≤650°C) dataset*

Figure 5 and Table 2 summarize the results of correlation analysis for the truncated dataset. Many physically meaningful features (i.e., volume fractions of phases and Ms) that we added into the raw 9Cr yield strength dataset commonly have high correlation coefficients. These highly impactful features from both PCC and MIC analyses are in good agreement with the generally accepted strengthening mechanisms, indicating that the features collected in the truncated dataset can capture the strengthening mechanisms of 9Cr steel well in the given temperature range. In this dataset, tensile testing temperature (TTTemp) is included, which allows its inclusion into the temperature-dependence of yield strength in the ML models. TTTemp possesses a strong negative correlation with yield strength, which is also consistent with the experimental observations that the higher the test temperature, the lower the yield strength.

There is a discrepancy between the results from MIC and PCC analyses, for example, MIC ranked wCo 1st (9th in PCC), while PCC ranked T2_VPV_M23C6 2nd (14th in MIC). This is attributed to the different algorithms in assigning in the strength of correlation. PCC only evaluates the strength of the linear relationship and MIC has an advantage over PCC when there is a non-linear correlation between input feature and target property. Detailed comparison of MIC and PCC analyses with different data structures are available in Ref.[37]. It should be emphasized that the purpose of performing both MIC and PCC analyses in this study is not to rank one method over the other. Correlation analysis is a topic of its own, aiming to study the statistical relationship strength between two variables. It is also a category of feature selection approach that facilitates the choice of the most relevant input features for ML[23]. The intent here is also to demonstrate that correlation analysis is necessary to validate whether or not underlying mechanisms have been efficiently captured by quantitatively evaluating the score of features considered. It can also be used to evaluate the quality of the consistency of a material dataset. The results of different correlation analyses can be further analyzed to inspire alloy design experts to generate alloy hypotheses.

[Figure 5 about here]

[Table 2 about here]

Five ML models (i.e., BR, LR, RF, NN, and SVM) were trained using the truncated dataset. The results are shown in Figure 6. Similarly, the number of top-ranking features based on the MIC and PCC analyses were varied to train these models. The accuracy of the models using the top-ranking features from the MIC and PCC analyses show similar trends. For example, increasing the top-ranking features from 5 to 10 for PCC, and from 5 to 15 for MIC increased the accuracy of these models significantly. After taking into account the top-ranking features, the accuracy of the BR, LR, RF, and SVM models was almost constant, with the NN model showing a monotonic decrease in accuracy. For the models utilized, it was necessary to include at least the top 10 features for PCC and the top 15 features for MIC to obtain good accuracy.

[Figure 6 about here]

Regardless of the type and number of features used for the PCC and MIC analyses, the accuracy of the trained models in predicting yield strength were, in order: RF>SVM>NN>BR≈LR. More specifically, RF, NN, and SVM exhibited very high accuracy ($R^2>0.9$), while the maximum accuracy of the LR and BR models were ~0.85. For example, Figure 7 shows the predicted yield strength using the RF model. It exhibits an excellent agreement with the experimentally determined yield strength. Although the accuracy of trained ML models with the dataset augmented by synthetic features is similar to those trained only with raw experimental data (see Figure 2), the fidelity of these models is notably enhanced for LR, BR, and SVM. This is because the synthetic features we incorporated into the dataset are proved to be highly correlated with the yield strength of 9Cr steel. Moreover, the ML models still achieve very high accuracy even though the truncated dataset contains ~10 % less data than the initial 9Cr dataset, mainly because the inconsistent data above 650°C was eliminated. As such, we believe that the trained ML models (as described in this section) are more accurate and can provide more realistic predictions.

[Figure 7 about here]

The high-fidelity surrogate models obtained in this work will allow prediction of the yield strength of hypothetical 9Cr alloys. However, in this case, additional work on predicting PAGS is required, as it was used as an input feature to predict the yield strength. For all features in groups 1 and 2 (see Figure 1), PAGS is unique. The PAGS is an essential input for predicting Ms[38], which was previously identified as a highly relevant feature for yield strength and served as an important constraint in training high-fidelity surrogate models. Also, PAGS depends on various details of the composition and processing conditions. However, PAGS of an alloy can only be obtained by physical inspection, i.e., metallography. Thus, following the similar workflow in the present study, surrogate models for PAGS were trained using the truncated dataset. The predicted PAGS using the NN, RF, and SVM models is in excellent agreement with experimental data (see Figure S2 in supplemental materials). As an example, a comparison between experimental and predicted PAGS of the 9Cr steel using the RF ML model is shown in Figure 8. We believe the outstanding performance of trained ML models is attributed to the extremely high correlation between input features and PAGS (see the correlation scores of high ranking features in Table S2). The average MIC score of top 15 features is 0.933±0.061, which is extremely high. The average scores of PCC are not as high as those of MIC, but the average score of top 10 is 0.660±0.100, which can be regarded to be high. With the success of this approach, PAGS for any 9Cr steel alloys can be derived and used as input to predict yield strength via a data analytics approach as demonstrated in the present study.

[Figure 8 about here]

In summary, we have demonstrated a workflow that can incorporate highly relevant physics into ML for predicting properties of complex heat-resistant alloys. Using a yield strength dataset of the 9-12 wt.% Cr steel as an example, the approach has been described in detail. We augmented raw experimental data with key features that can capture both the microstructure and phase transformation of this class of alloy, i.e., the volume fraction of key phases, A3, and martensite phase transformation temperatures. It is worth mentioning that the present features could not capture the complex location- and size-specific microstructural detail of the secondary phases that form in the 9Cr alloys. It would be ideal to incorporate such detailed microstructure-related information into the data analytics workflow. However, obtaining such a large volume of high-fidelity microstructural details for all the alloy chemistries and processing conditions will be extremely time and cost-prohibitive.

We computed these synthetic features using high-fidelity thermodynamic models in a high-throughput manner. Critical evaluation of each temperature-based sub-datasets, including correlation analysis and ML training, showed that data above 650°C are insufficient for correctly capturing the significant factors related to the yield strength of 9Cr steel due to the relative lack of experimental data and relevant microstructure features. Thus, this information was removed from the 9Cr dataset, and correlation analysis of this truncated dataset showed that the high-ranking features were in good agreement with the generally accepted strengthening mechanisms.

We tested the performance of representative ML models, i.e., RF, SVM, NN, BR, and LR, as a function of the number of top-ranking features. From this exercise, the top 10 features from PCC and the top 15 features from MIC are necessary to obtain good accuracy for all models. Among the ML models tested, the RF and SVM ones exhibited very high accuracy ($R^2>0.95$) for predicting 9Cr steel yield strength. In conclusion, this study demonstrated that high-fidelity surrogate models could be trained with highly relevant and physically meaningful features. Such physical constraints effectively prevent erroneously predicting properties of hypothetical candidate alloys when interrogating trained ML models in a data-driven materials design. We anticipate that the approach demonstrated in the present work can be further extended by integrating additional alloy physical/chemical features beyond what is achievable in this study.

**Methods**

*Experimental dataset and synthetic alloy features via thermodynamic calculations*

The raw experimental dataset was compiled by National Energy Technology Laboratory[8,9], USA, using the creep datasheet for high Cr steel[39] in the MatNavi materials database by the National Institute for Materials Science, Japan. The dataset is consists of compositions of 18 elements, processing and testing temperatures, and prior austenite grain size (PAGS, converted from austenite grain size number). The state-of-the-art steel and Fe-alloys dataset TCFE9 [40] was used to compute the volume fractions of the phases and the A3 temperature for each steel composition by the CALPHAD approach[41]. A recently developed thermodynamic model[38] (also implemented in Thermo-Calc software package[42,43]) was adopted to calculate martensite start temperatures (Ms). This analytical model, which is an extension of the models developed by Borgenstam and Hillert[44], and Stormvinter et al.[45], takes into account of the thermodynamic driving force for of FCC-BCC phase transformation as the major contribution as well as PAGS as a non-chemical contribution to predict Ms of a given 9Cr alloy. Raw experimental data were augmented with these synthetic features by the high-throughput calculation using Thermo-Calc, resulting in a dataset with 451 instances/rows, 45 input features/columns, and one target (0.2% yield strength), and the temperature range of room temperature (RT) to 800°C.

*Correlation analysis*

The necessity of correlation analysis in materials data analytics is threefold: (1) validate if high-ranking features are consistent with generally accepted mechanisms; (2) provide a numerical/statistical basis for the selection of input features in the training of ML models; and (3) facilitate the generation of alloy hypotheses by identifying overlooked/hidden features in previous work. The correlation between the input features and the target was represented by Pearson's correlation coefficient (PCC)[34] and maximal information coefficient (MIC)[35]. While PCC only evaluates the strength of the linear relationship, MIC identifies the strength of both linear and nonlinear relationships. The correlation coefficient of PCC lies between -1 and 1, where 1 indicates a total positive linear correlation, -1 indicates a complete negative/reciprocal linear correlation, and 0 indicates no linear correlation. The closer the coefficient is to 1 or -1, the stronger the correlation between the two variables is. The correlation coefficient of MIC ranges between 0 and 1. The closer the coefficient is to 1, the stronger is that the correlation.

*Machine learning*

The performance of five representative ML models was studied: (1) linear regression (LR)[28], (2) Bayesian ridge (BR)[29,30], (3) *k*-nearest neighbor (NN)[31], (4) random forest (RF)[32], and (5) support vector machines (SVM)[33]. A different number of top-ranking features based on the ranking from MIC and |PCC| (i.e., the absolute value of the correlation coefficient of PCC) was used to train ML models and evaluate their performance. The hyperparameters of each model were tuned by using up to 150 iterations to identify the optimum parameters. Each model was trained ten times for a given set of features to determine the averaged accuracy and its standard deviation. The ranking from correlation analysis does not assign any hierarchical factor to the features, i.e., all features have the same weight in ML training regardless of their ranking. The coefficient of determination ($R^2$) was adopted to represent the accuracy of ML models. The correlation analysis and ML were performed using the open-source data analytics toolkit, **A**dvanced data **SCiE**nce toolkit for **N**on-**D**ata **S**cientists (ASCENDS)[46,47], which is available via GitHub (https://github.com/ornlpmcp/ASCENDS).

Depending on the flexibility of the ML models, overfitting or underfitting the data is possible. The *k*-fold approach[48] with *k* = 5 was used for the ML training. Four groups were used to train the machine learning model, and the one remaining group (i.e., unseen data) was withheld during training and later used as the validation data to evaluate the accuracy of models. Then we have trained the same ML model (i.e., the same feature set for a given ML algorithm) ten times to get the statistics for uncertainty quantification. As such, it ensured that the fitting of the ML models to the data was balanced.

**Acknowledgments**

This research was sponsored by the U.S. Department of Energy, Office of Fossil Energy, eXtreme environment MATerials (XMAT) consortium. This research used resources of the Compute and Data Environment for Science (CADES) at the Oak Ridge National Laboratory, which is supported by the Office of Science of the U.S. Department of Energy under Contract No. DE-AC05-00OR22725. The authors thank YiYu Wang for valuable discussion and Chris Layton for his support on using CADES.

**Data availability**

The data that support the findings of this study are available from the corresponding authors upon reasonable request.

**Author contributions**

D.S. conceived the study. J.A.H provided the dataset. J.P. performed correlation analysis and machine learning training. J.P., Y.Y, and D.S. analyzed the data. J.P. drafted the manuscript. Y.Y., J.A.H., E.L-C., and D.S. reviewed the manuscript.

**Competing interests**

The authors declare no competing interests.

**List of Tables**

Table 1 List of alloy features considered in this work to predict 0.2% yield strength (MPa) of 9-12wt% Cr steels

Table 2 Top 20 features from the correlation analysis between alloy features (simple features plus synthetic features populated from the high-throughput calculation) and yield strength using the MIC and PCC methods for the truncated (≤650°C) dataset. Features from the PCC analysis with a negative impact on yield strength are presented in parenthesis. The corresponding correlation coefficients are reported in Figure 5.

Table 1 List of alloy features considered in this work to predict 0.2% yield strength (MPa) of 9-12wt% Cr steels

| Features | | Descriptions |
|---|---|---|
| **Compositions, Processing and test conditions** (Simple features /raw data) | wElements | Elemental composition (wt. %) |
| | Normal (T1) | Normalization temperature |
| | Temper 1 (T2) | Tempering temperature 1 (°C) |
| | Temper 2 | Tempering temperature 2 (°C) |
| | TTTemp | Testing temperature (°C) (RT, 100, 200, 300, 400, 450, 500, 550, 600, 650, 700, 750, 800°C) |
| | PAGS | Prior austenite grain size (μm) |
| **Microstructure, Phase transformation temperatures** (Synthetic features) | Temperature_VPV_Phases | Volume fraction of phases at T1 and T2 |
| | A3 | The temperature at which transformation of ferrite to austenite is completed during heating (°C) |
| | Ms | Martensite start temperature (°C) |
| | dT | Difference between T1 and A3 (°C) |

Table 2 Top 20 features from the correlation analysis between alloy features (simple features plus synthetic features populated from the high-throughput calculation) and yield strength using the MIC and PCC methods for the truncated (≤650°C) dataset. Features from the PCC analysis with a negative impact on yield strength are presented in parenthesis. The corresponding correlation coefficients are reported in Figure 5.

| Features | MIC ranking | |PCC| ranking | Features | |PCC| ranking | MIC ranking |
|---|---|---|---|---|---|
| wCo | 1 | 9 | wNi | 1 | 12 |
| TTTemp | 2 | (4) | T2_VPV_M23C6 | 2 | 14 |
| wCr | 3 | 10 | Ms | (3) | 8 |
| wS | 4 | (35) | TTTemp | (4) | 2 |
| T2_VPV_MNS | 5 | (34) | wC | 5 | 11 |
| T1_VPV_MNS | 6 | (31) | wFe | (6) | 7 |
| wFe | 7 | (6) | Temper1 | (7) | 10 |
| Ms | 8 | (3) | wV | 8 | 15 |
| wNb | 9 | (41) | wCo | 9 | 1 |
| Temper1 | 10 | (7) | wCr | 10 | 3 |
| wC | 11 | 5 | Normal | 11 | 22 |
| wNi | 12 | 1 | PAGS | 12 | 18 |
| wW | 13 | 14 | wMn | 13 | 19 |
| T2_VPV_M23C6 | 14 | 2 | wW | 14 | 13 |
| wV | 15 | 8 | dT | 15 | 26 |
| wN | 16 | 32 | T2_VPV_FCC | (16) | 39 |
| T2_VPV_Z_PHASE | 17 | 20 | wAl | 17 | 30 |
| PAGS | 18 | 12 | T1_VPV_M23C6 | 18 | 35 |
| wMn | 19 | 13 | T2_VPV_BCC_FE | 19 | 21 |
| T2_VPV_VN | 20 | (40) | T2_VPV_Z_PHASE | 20 | 17 |

**List of Figures**

Figure 1 Alloy features considered in the 9Cr data analytics. Groups 1/2 are raw experiment data, and groups 3/4 are computed synthetic alloy features. This dataset cover data from room temperature to 800°C and the values in parentheses next to temperatures represent the number of data points at each temperature. Complete details of this dataset are reported in Table 1.

Figure 2 Accuracy of five trained ML models (BR: Bayesian ridge regression, LR: linear regression, NN: nearest neighbor, RF: random forest, and SVM: support vector machines) with raw experimental data (compositions, processing, and testing conditions, and PAGS) as a function of the number of top-ranking features in the entire 9Cr dataset. The hyperparameters of each model were tuned up to 150 iterations to obtain optimum parameters. Each model was trained 10 times to determine the average accuracy and its standard deviation.

Figure 3 Results of correlation analysis between all features (composition, processing and test conditions, microstructure, and phase transformation temperature, see Table 1) and yield strength at selected representative temperatures. The correlation coefficients of the top 10 and bottom 10 features out of 45 features from 200°C (low temperature), 550 and 650°C (medium to high temperatures), and 750°C (above service temperature) are presented.

Figure 4 Accuracy of trained ML models (random forest as an example) as a function of temperature and numbers of top-ranking features from the (a) MIC and (b) PCC analyses. The models were trained for 10 times to determine the average accuracy and its standard deviation (error bar). The hyperparameters of each model were tuned up to 150 iterations to obtain optimum parameters. The vertical dash line indicates where issues of lack of data and relevant features start to occur.

Figure 5 Correlation analysis between all alloy features and yield strength in the truncated (≤650°C) dataset. (a) top 10 and bottom 10 features from the PCC analysis, and (b) top 20 ranking features from MIC and the corresponding features from |PCC|. The ranking of each feature from both analyses is reported in Table 2.

Figure 6 Accuracy of five trained ML models (BR: Bayesian ridge regression, LR: linear regression, NN: nearest neighbor, RF: random forest, and SVM: support vector machines regression) in predicting yield strength. These models were trained with synthetic features populated from high-throughput calculation as a function of the number of top-ranking features in the truncated (≤650°C) dataset. The hyperparameters of each model were tuned up to 150 iterations to obtain optimum parameters. Each model was trained 10 times to determine the average accuracy and its standard deviation (error bar).

Figure 7 Experimental vs. predicted yield strength of the 9Cr steel with random forest (RF) with the top 10 features from both PCC and MIC analysis. MAE stands for mean absolute error (MAE).

Figure 8 Experimental vs. predicted PAGS of the 9Cr steel with random forest (RF) with the top 10 features from MIC and PCC analyses. MAE stands for mean absolute error (MAE).

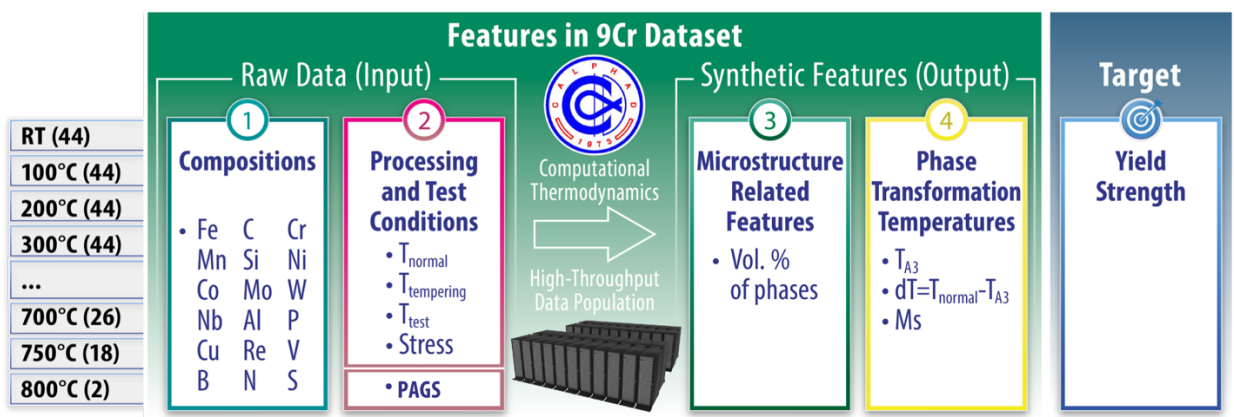

Figure 1 Alloy features considered in the 9Cr data analytics. Groups 1/2 are raw experiment data, and groups 3/4 are computed synthetic alloy features. This dataset cover data from room temperature to 800°C and the values in parentheses next to temperatures represent the number of data points at each temperature. Complete details of this dataset are reported in Table 1.

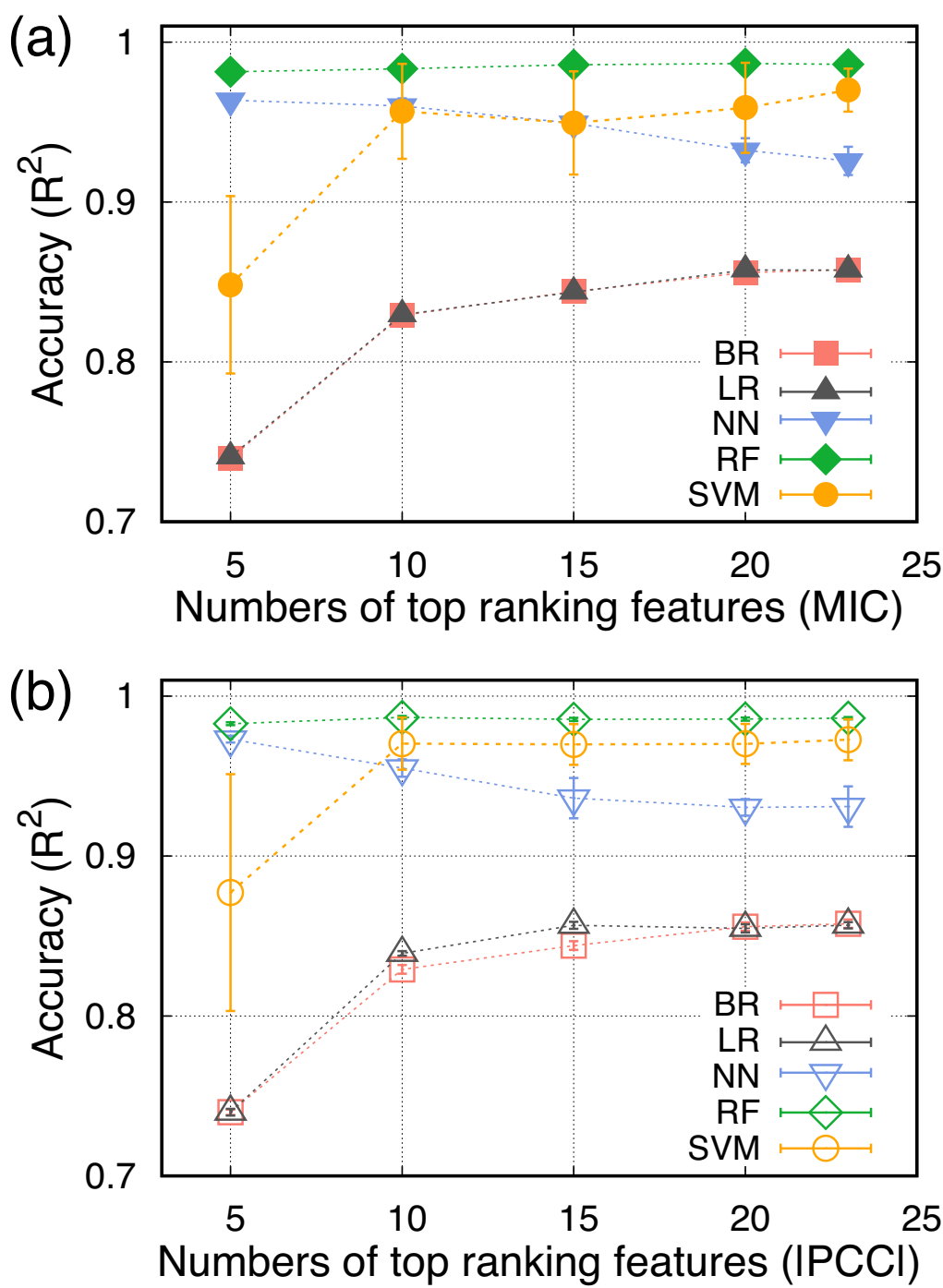

Figure 2 Accuracy of five trained ML models (BR: Bayesian ridge regression, LR: linear regression, NN: nearest neighbor, RF: random forest, and SVM: support vector machines) with raw experimental data (compositions, processing, and testing conditions, and PAGS) as a function of the number of top-ranking features in the entire 9Cr dataset. The hyperparameters of each model were tuned up to 150 iterations to obtain optimum parameters. Each model was trained 10 times to determine the average accuracy and its standard deviation.

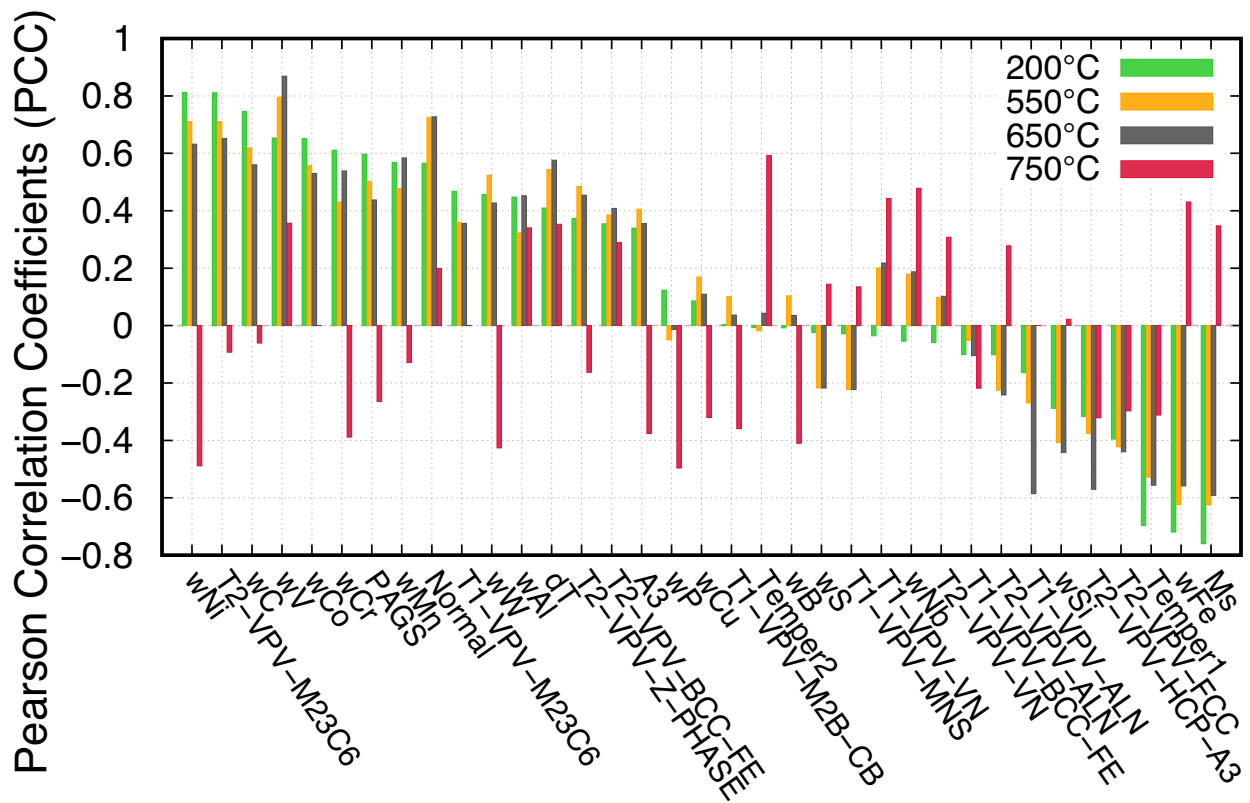

Figure 3 Results of correlation analysis between all features (composition, processing and test conditions, microstructure, and phase transformation temperature, see Table 1) and yield strength at selected representative temperatures. The correlation coefficients of the top 10 and bottom 10 features out of 45 features from 200°C (low temperature), 550 and 650°C (medium to high temperatures), and 750°C (above service temperature) are presented.

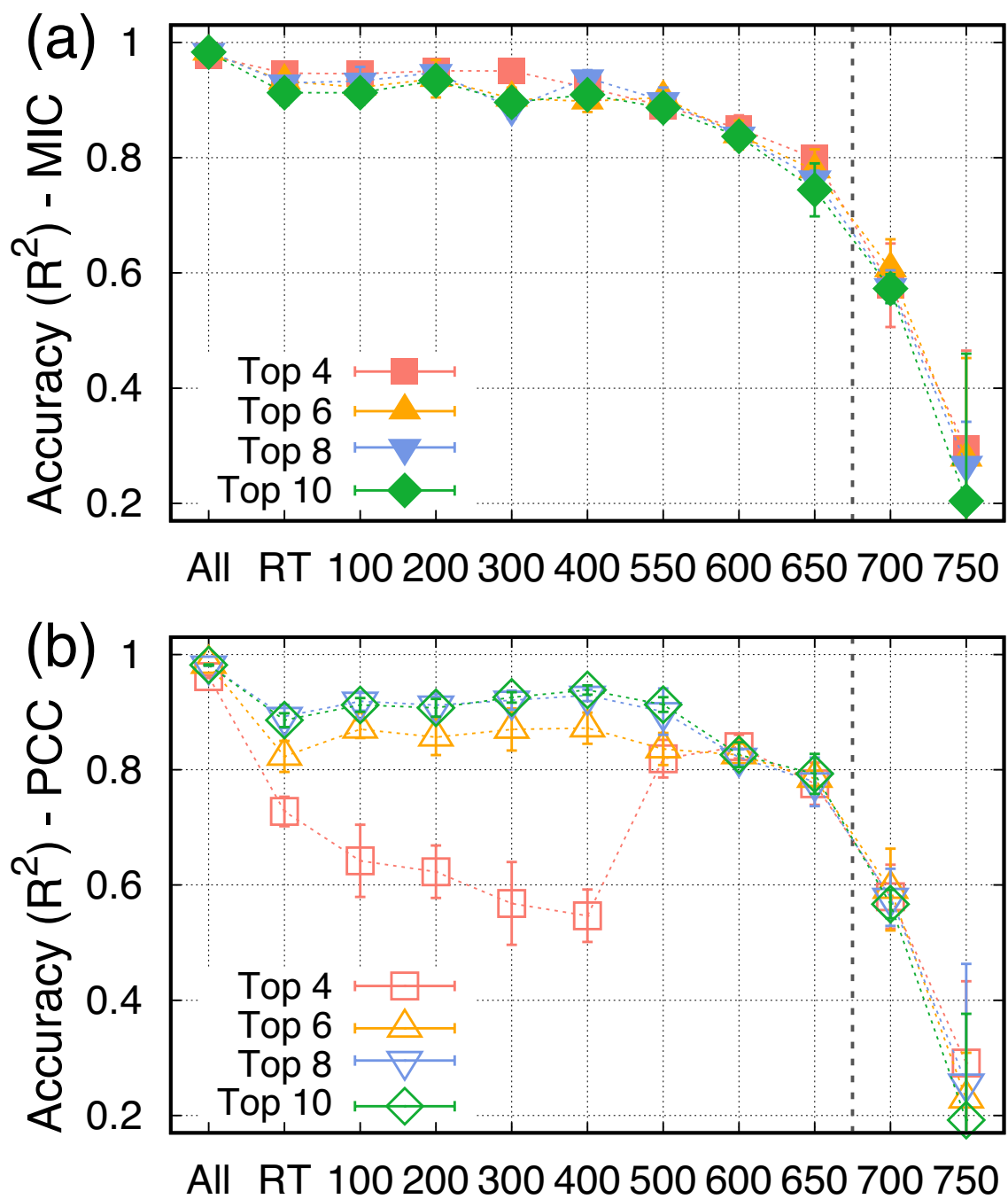

Figure 4 Accuracy of trained ML models (random forest as an example) as a function of temperature and numbers of top-ranking features from the (a) MIC and (b) PCC analyses. The models were trained for 10 times to determine the average accuracy and its standard deviation (error bar). The hyperparameters of each model were tuned up to 150 iterations to obtain optimum parameters. The vertical dash line indicates where issues of lack of data and relevant features start to occur.

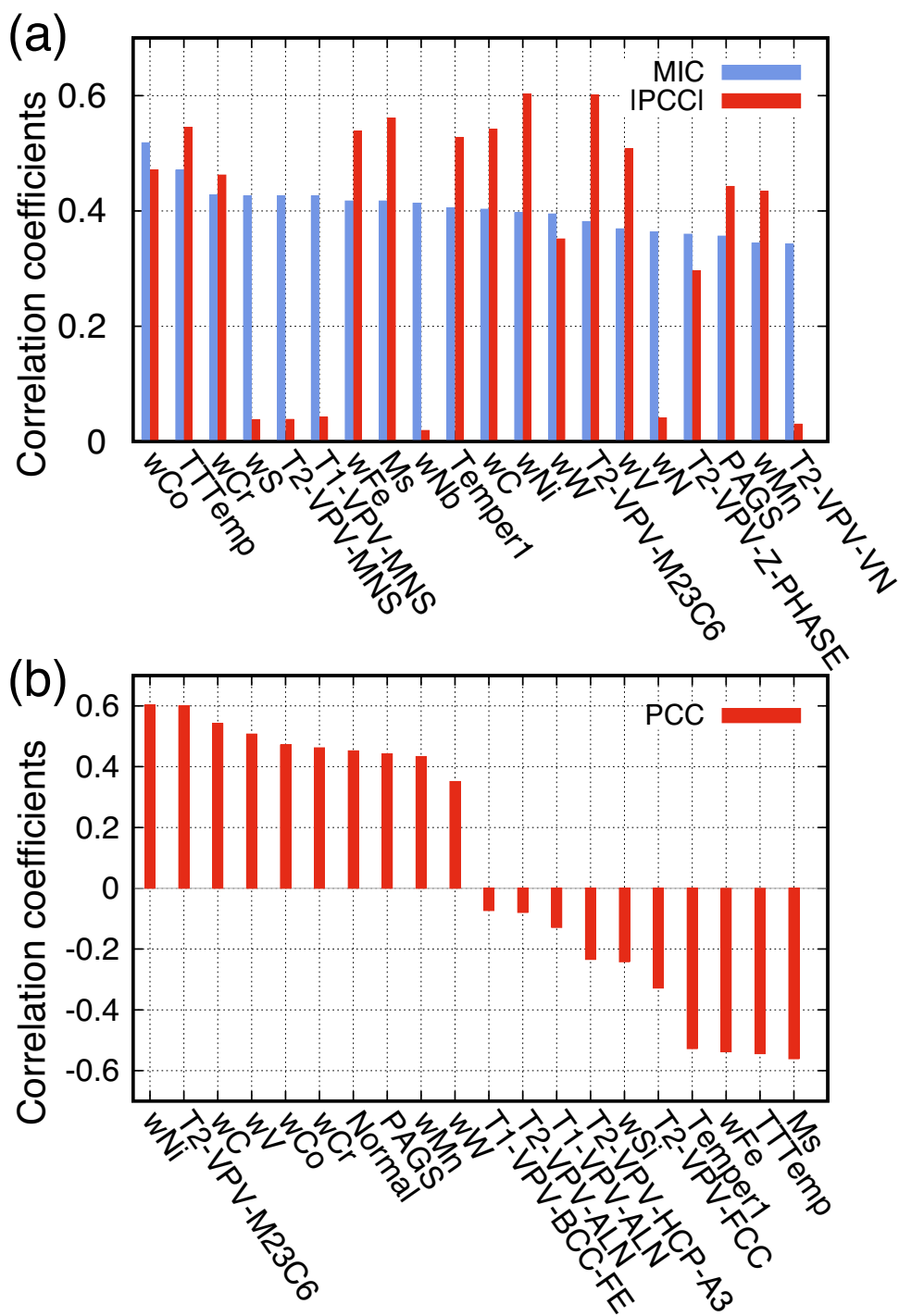

Figure 5 Correlation analysis between all alloy features and yield strength in the truncated (≤650°C) dataset. (a) top 10 and bottom 10 features from the PCC analysis, and (b) top 20 ranking features from MIC and the corresponding features from |PCC|. The ranking of each feature from both analyses is reported in Table 2.

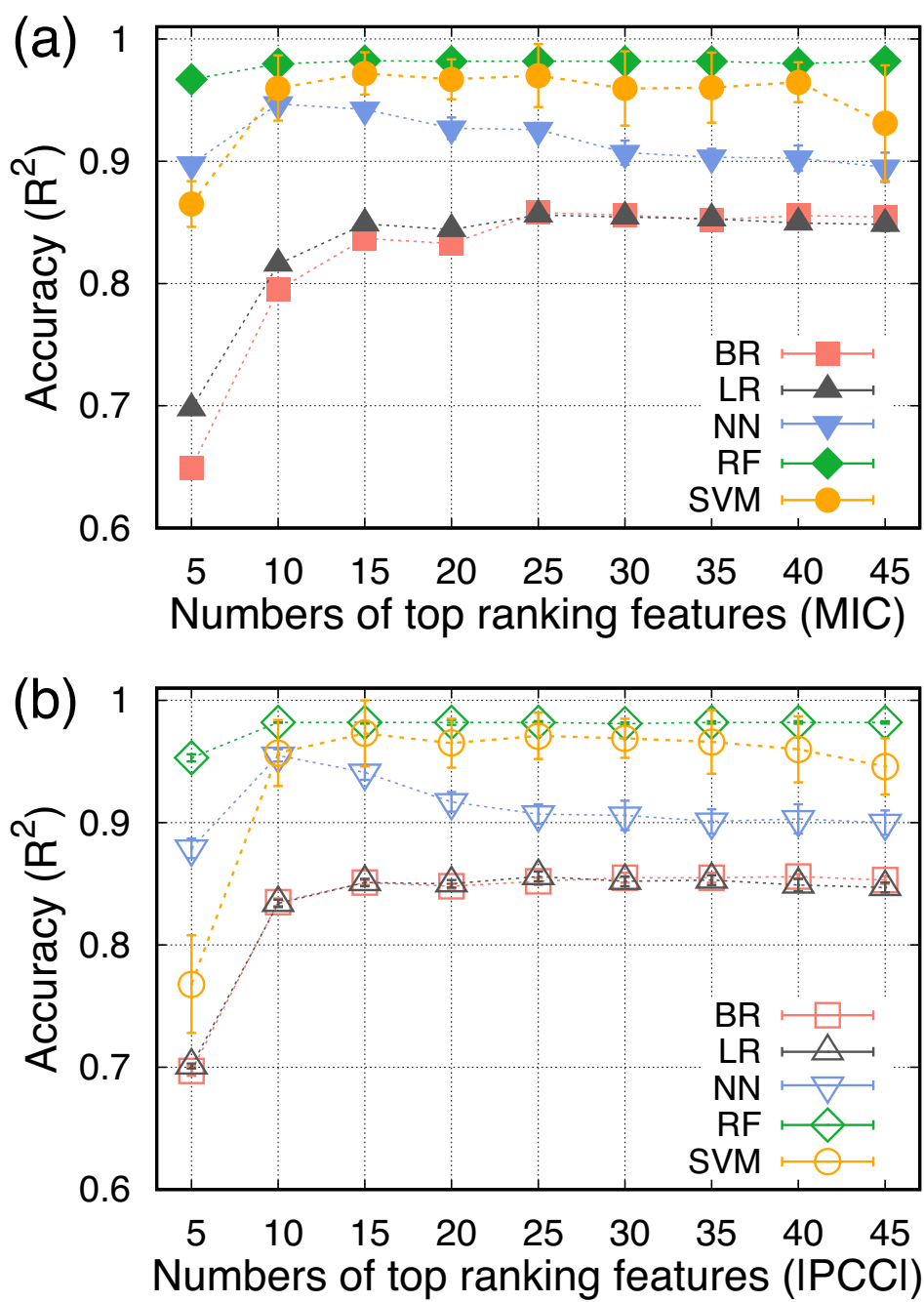

Figure 6 Accuracy of five trained ML models (BR: Bayesian ridge regression, LR: linear regression, NN: nearest neighbor, RF: random forest, and SVM: support vector machines regression) in predicting yield strength. These models were trained with synthetic features populated from high-throughput calculation as a function of the number of top-ranking features in the truncated (≤650°C) dataset. The hyperparameters of each model were tuned up to 150 iterations to obtain optimum parameters. Each model was trained 10 times to determine the average accuracy and its standard deviation (error bar).

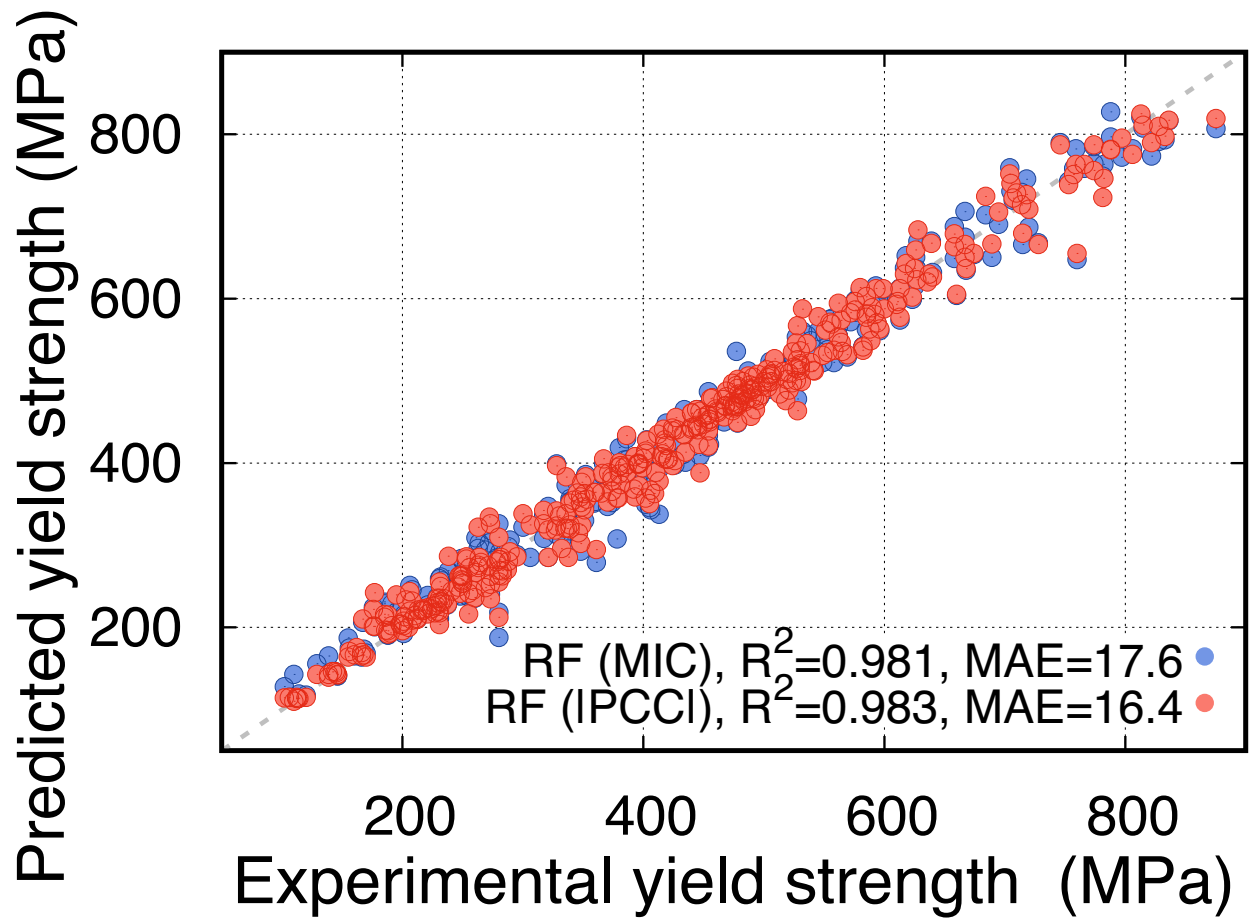

Figure 7 Experimental vs. predicted yield strength of the 9Cr steel with random forest (RF) with the top 10 features from both PCC and MIC analysis. MAE stands for mean absolute error (MAE).

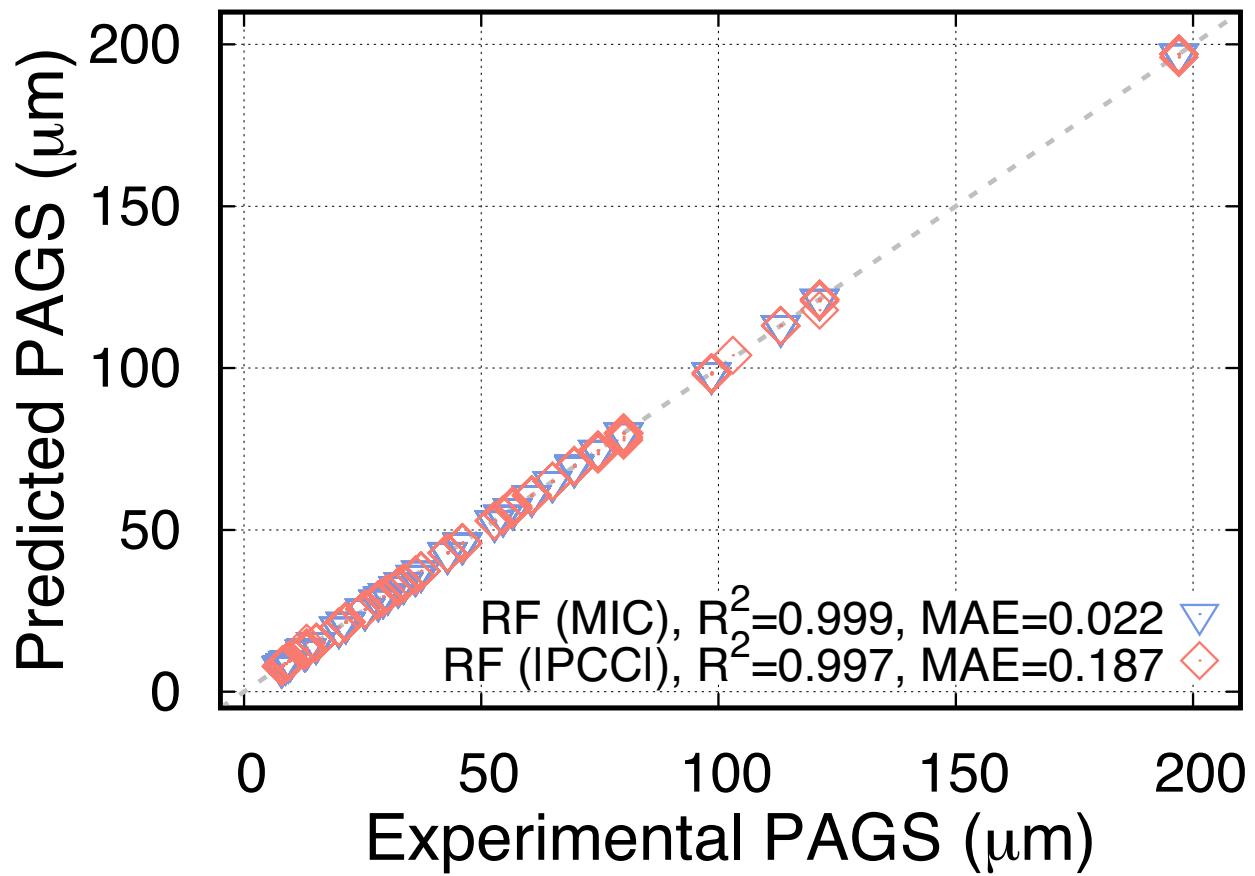

Figure 8 Experimental vs. predicted PAGS of the 9Cr steel with random forest (RF) with the top 10 features from MIC and PCC analyses. MAE stands for mean absolute error (MAE).